\documentclass[12pt,english]{article}
\usepackage[T1]{fontenc}
\usepackage[latin9]{inputenc}
\usepackage{geometry}
\usepackage{babel}
\usepackage{amstext}
\usepackage{graphicx}
\usepackage{setspace}
\usepackage[unicode=true,pdfusetitle,
 bookmarks=true,bookmarksnumbered=false,bookmarksopen=false,
 breaklinks=false,pdfborder={0 0 1},backref=false,colorlinks=false]
 {hyperref}
\usepackage{breakurl}

\makeatletter
\usepackage{amsthm}
\usepackage{amssymb}
\usepackage{cite}

\makeatother

\begin{document}
~

\vspace{10mm}

\noindent \begin{center}
{\LARGE{}Bose-Einstein condensation of photons and grand-canonical
condensate fluctuations}
\par\end{center}{\LARGE \par}

\noindent \begin{center}
Jan Klaers\vspace{-6mm}

\par\end{center}

\noindent \begin{center}
\textit{Institute for Applied Physics, University of Bonn, Germany}\vspace{-6mm}

\par\end{center}

\noindent \begin{center}
\textit{Present address:}\vspace{-6mm}

\par\end{center}

\noindent \begin{center}
\textit{Institute for Quantum Electronics, ETH Z\"urich, Switzerland}
\par\end{center}

\noindent \begin{center}
Martin Weitz\vspace{-6mm}

\par\end{center}

\noindent \begin{center}
\textit{Institute for Applied Physics, University of Bonn, Germany}\vspace{3mm}

\par\end{center}
\begin{abstract}
We review recent experiments on the Bose-Einstein condensation of
photons in a dye-filled optical microresonator. The most well-known
example of a photon gas, photons in blackbody radiation, does not
show Bose-Einstein condensation. Instead of massively populating the
cavity ground mode, photons vanish in the cavity walls when they are
cooled down. The situation is different in an ultrashort optical cavity
imprinting a low-frequency cutoff on the photon energy spectrum that
is well above the thermal energy. The latter allows for a thermalization
process in which both temperature and photon number can be tuned independently
of each other or, correspondingly, for a non-vanishing photon chemical
potential. We here describe experiments demonstrating the fluorescence-induced
thermalization and Bose-Einstein condensation of a two-dimensional
photon gas in the dye microcavity. Moreover, recent measurements on
the photon statistics of the condensate, showing Bose-Einstein condensation
in the grandcanonical ensemble limit, will be reviewed.\vspace{3mm}

\end{abstract}

\section{Introduction}

Quantum statistical effects become relevant when a gas of particles
is cooled, or its density is increased, to the point where the associated
de Broglie wavepackets spatially overlap. For particles with integer
spin (bosons), the phenomenon of Bose-Einstein condensation (BEC)
then leads to macroscopic occupation of a single quantum state at
finite temperatures \cite{leggett2006}. Bose-Einstein condensation
in the gaseous case was first achieved in 1995 by laser and subsequent
evaporative cooling of a dilute cloud of alkali atoms \cite{Anderson:1995,Davis:1995p2059,Bradley:1997},
as detailed in preceding chapters of this volume. The condensate atoms
can be described by a macroscopic single-particle wavefunction, similar
as known from liquid helium \cite{leggett2006}. Bose-Einstein condensation
has also been observed for exciton-polaritons, which are hybrid states
of matter and light \cite{deng2002,Kasprzak:2006p443,Balili:2007p1342},
magnons \cite{Demokritov:2006p1354}, and other physical systems,
see the following chapters of this volume. Other than material particles,
photons usually do not show Bose-Einstein condensation \cite{Huang:StatisticalMechanics1987}.
In blackbody radiation, the most common Bose gas, photons at low temperature
disappear, instead of condensing to a macroscopically occupied ground
state mode. In this system, photons have a vanishing chemical potential,
meaning that the number of photons is determined by the available
thermal energy and cannot be tuned independently from temperature.
Clearly, a precondition for a Bose-Einstein condensation of photons
is a thermalization process that allows for an independent adjustment
of both photon number and temperature. Early theoretical work has
proposed a thermalization mechanism by Compton scattering in plasmas
\cite{Zeldovich:1969p1287}. Chiao et al. proposed a two-dimensional
photonic quantum fluid in a nonlinear resonator \cite{Chiao:OpticsCommunications2000}.
Thermal equilibrium here was sought from photon-photon scattering,
in analogy to atom-atom scattering in cold atom experiments, but the
limited photon-photon interaction in available nonlinear materials
has yet prevented a thermalization \cite{Mitchell:2000p1910}. In
the strong coupling regime, (quasi-)equilibrium Bose-Einstein condensation
of exciton-polaritons, mixed states of matter and light, has been
achieved \cite{deng2002,Kasprzak:2006p443,Balili:2007p1342}. Here
interparticle interactions of the excitons drive the system into or
near thermal equilibrium. More recently, evidence for superfluidity
of polaritons has been reported \cite{amo2009,Amo:NaturePhysics2009}.
Other experimental work has observed the kinetics of condensation
of classical optical waves \cite{sun2012}.

\begin{figure}
\begin{centering}
\includegraphics[width=0.55\paperwidth]{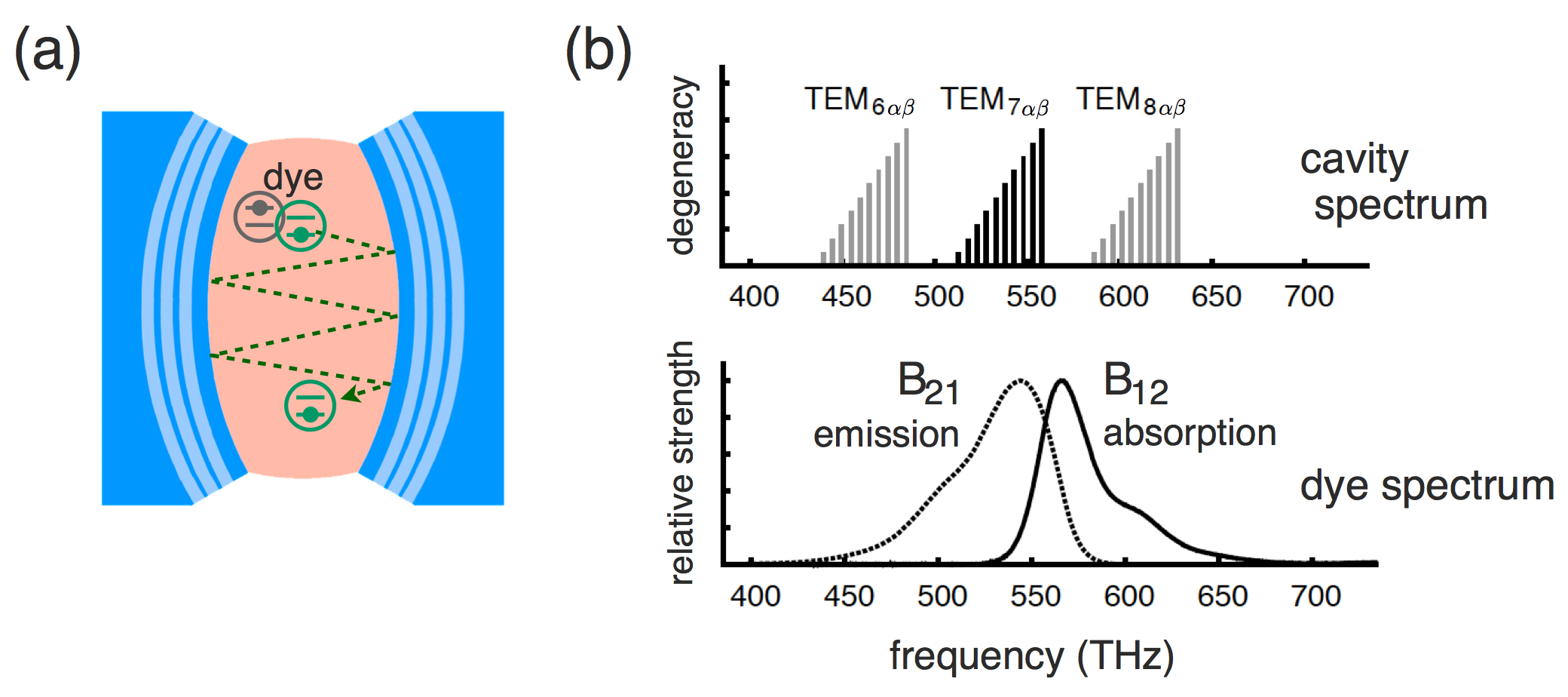}
\par\end{centering}

\centering{}%
\begin{minipage}[t]{0.93\columnwidth}%
\textbf{\small{}Figure 1 |}{\small{} \label{fig:Figure1}}\textbf{\small{}(a)}{\small{}
Scheme of the experimental setup used in \cite{Klaers:2010}. }\textbf{\small{}(b)}{\small{}
Schematic spectrum of cavity modes. Transverse modes belonging to
the manifold of longitudinal mode number $q=7$ are shown by black
lines, those of other longitudinal mode numbers in grey. The bottom
graph indicates the relative absorption coefficient and fluorescence
strength of rhodamine 6G dye versus frequency.}%
\end{minipage}
\end{figure}

Bose-Einstein condensation of photons in a dye filled microresonator
has been realized in 2010 in our group at the University of Bonn and
in 2014 at Imperial College London \cite{Klaers:2010,marelic2014}.
Thermalization of the photon gas with the dye solution is achieved
by repeated absorption and re-emission processes. For liquid dye solutions
at room temperature conditions, it is known that rapid decoherence
from frequent collisions ($10^{-14}\,\textrm{s}$ timescale) with
solvent molecules prevent a coherent excitation exchange between photonic
and electronic degrees of freedom \cite{DeAngelis:2000p484,Yokoyama:1989p2104},
so that the condition of strong light-matter coupling is not met.
It is therefore justified to regard the bare photonic and electronic
excitations of the system as the true energy eigenstates. The separation
between the two curved resonator mirrors, see Fig. \ref{fig:Figure1}a,
is of order of the photon wavelength.
 The small cavity spacing causes a large frequency spacing between
the longitudinal resonator modes, which is of order of the emission
width of the dye molecules, see Fig. \ref{fig:Figure1}b. Under these
conditions, only photons of a fixed longitudinal mode are observed
to populate the resonator, which effectively makes the photon gas
two-dimensional as only the two transversal motional degrees of freedom
remain. The lowest lying mode of this manifold ($q=7$), the $\textrm{TEM}_{00}$
transverse ground mode, acts as a low frequency cutoff at an energy
of $\hbar\omega_{\textrm{c}}\simeq2.1\,\textrm{eV}$ in the Bonn experiment
\cite{Klaers:2010,Klaers:2010p2137}. This restricts the photon spectrum
to energies $\hbar\omega$ well above the thermal energy $k_{\text{B}}T\simeq1/40\,\textrm{eV}$,
i.e. $\hbar\omega\ge\hbar\omega_{\textrm{c}}\gg k_{\text{B}}T$, which
to good approximation decouples the number of photons from the heat
content of the system (non-vanishing chemical potential). In this
situation, the photon number becomes tunable by (initial) optical
pumping, which can be regarded as fully analogous to the loading of
cold atoms into a magnetic or optical dipole trap.

In the microcavity, the energy-momentum relation moreover becomes
quadratic, as for a massive particle, and the mirror curvature induces
an effective trapping potential in the transverse plane. In general,
significant population of high transverse modes ($\textrm{TEM}_{\alpha\beta}$
with high transversal mode numbers $\alpha$ and $\beta$ and, correspondingly,
high eigenfrequencies) are expected at high temperatures, while the
population concentrates to the lowest transverse modes, when the system
is cold. One can show that the photon gas in the resonator is formally
equivalent to a harmonically trapped two-dimensional gas of massive
bosons with effective mass $m_{\textrm{ph}}=\hbar\omega_{\textrm{c}}(n/c)^{2}$.
Here $c$ denotes the vacuum speed of light and $n$ the index of
refraction of the resonator medium. In thermal equilibrium, such a
system is known to undergo Bose-Einstein condensation at a finite
temperature \cite{Bagnato:1991p552}. Both the thermalization of the
photon gas to room temperature \cite{Klaers:2010p2137} and the Bose-Einstein
condensation \cite{Klaers:2010} has been verified experimentally.

Several theoretical publications have discussed different aspects
of photon Bose-Einstein condensation using a variety of approaches
\cite{Klaers2:2011,Sobyanin:2012,Kirton:2013,snoke2013,deLeeuw:2013,nyman2014,kruchkov2014,strinati2014,deLeeuw2014,sela2014,vanderWurff2014},
including work based on a superstatistical approach \cite{Sobyanin:2012},
on a Schwinger-Keldysh theory \cite{deLeeuw:2013}, and on a master
equation approach \cite{Kirton:2013}. Investigated topics include
first-order coherence properties such as the dynamics of phase coherence
onset \cite{snoke2013}, and equilibrium phase fluctuations of the
photon condensate \cite{deLeeuw2014}. Moreover, second-order coherence
properties of photon condensates have been studied in some detail
\cite{Klaers2:2011,Sobyanin:2012,vanderWurff2014}. The coupling of
the photon gas to the dye medium, which allows for both energy and
particle exchange, can be described by a grand-canonical ensemble
representation. This leads to physically observable consequences in
the condensed phase regime, in which the condensate performs anomalously
large intensity fluctuations \cite{Klaers2:2011}. Another key topic
is the relation between lasing and condensation. These different regimes
have been studied in a theory model accounting for photon loss leading
to partial thermal equilibrium of the photon gas \cite{Kirton:2013}.

In the following, section 2 gives a theoretical description of the
fluorescence induced thermalization mechanism, as well as the expected
thermodynamic behavior of the two-dimensional photon gas in the dye-filled
microcavity system. Further, section 3 describes experiments observing
the thermalization and Bose-Einstein condensation of the photon gas
at room temperature. Section 4 reviews theory and experimental results
regarding the grand-canonical nature of the condensate fluctuations.
Finally, section 5 concludes this contribution.

\section{Thermodynamics of a two-dimensional photon gas}

\subsection{Thermal and chemical equilibrium}

In the dye-filled microcavity system, the photon gas in the resonator
is thermally coupled to the dye medium. This thermalization mechanism
relies on two pre-conditions. First, the dye medium itself has to
be in thermal equilibrium. Consider an idealized dye molecule with
an electronic ground state and an electronically excited state separated
by the energy $\hbar\omega_{\textrm{ZPL}}$ (zero-phonon-line), each
subject to additional rotational and vibrational level splitting \cite{Lakowicz:1999}.
Frequent collisions of solvent molecules with the dye, on the timescale
of a few femtoseconds at room temperature, rapidly alter the rovibrational
state of the dye molecules. These collisions are many orders of magnitude
faster than the electronic processes (the upper state natural lifetime
of e.g. rhodamine 6G dye is $4\,\textrm{ns}$), so that both absorption
and emission processes will take place from an equilibrated internal
state. One can show that the Einstein coefficients for absorption
and emission $B_{12,21}(\text{\ensuremath{\omega}})$ then will be
linked by a Boltzmann factor
\begin{equation}
\frac{B_{21}(\omega)}{B_{12}(\omega)}=\frac{w_{\downarrow}}{w_{\uparrow}}\,e^{-\frac{\hbar(\omega-\omega_{\text{ZPL}})}{k_{\textrm{B}}T}}\;\textrm{,}\label{eq:KS}
\end{equation}
where $w_{\downarrow,\uparrow}$ are statistical weights related to
the rovibrational density of states \cite{Klaers2:2011}. This relation
is known as the Kennard-Stepanov law \cite{Kennard:1918p1291,Kennard:1927p1292,Stepanov:1957,Stepanov:1957_2,McCumber:1964p1463,Lakowicz:1999}
. Experimentally, the Kennard-Stepanov relation is well fulfilled
for many dye molecules. Deviations from this law can either arise
from imperfect rovibrational relaxation or a reduced fluorescence
quantum yield \cite{klaers2014}. 

The second pre-condition for the light-matter thermalization process
is the chemical equilibrium between photon gas and dye medium. Absorption
and emission processes can be regarded as a photochemical reaction
of the type $\gamma+\downarrow\:\leftrightharpoons\:\uparrow$ between
photons ($\gamma$), excited ($\uparrow$) and ground state ($\downarrow$)
molecules. This reaction reaches chemical equilibrium, if the rates
of competing processes (such as pump and loss) are negligible and
there is no net change in the densities of one of the species anymore.
The corresponding chemical potentials then satisfy $\mu_{\gamma}+\mu_{\downarrow}=\mu_{\uparrow}$,
which can also be expressed as \cite{Klaers2:2011}
\begin{equation}
e^{\frac{\mu_{\gamma}}{k_{\text{B}}T}}=\frac{w_{\downarrow}}{w_{\uparrow}}\,\frac{\rho_{\uparrow}}{\rho_{\downarrow}}\,e^{\frac{\hbar\omega_{\text{ZPL}}}{k_{B}T}}\;\textrm{,}\label{eq:fugacity}
\end{equation}
where $\rho_{\uparrow}$ ($\rho_{\downarrow}$) denotes the density
of excited (ground state) molecules. In equilibrium, the photon chemical
potential is thus determined by the excitation ratio $\rho_{\uparrow}/\rho_{\downarrow}$
of the medium. Assuming both the Kennard-Stepanov law eq. (1) and
chemical equilibrium, as expressed by eq. (2), one can show that multiple
absorption-emission cycles drive the photon gas into thermal equilibrium
with the dye solution at temperature $T$, and with a photon chemical
potential $\mu_{\gamma}$ determined by the molecular excitation ratio
\cite{Klaers2:2011}.

\subsection{Cavity photon dispersion and BEC criticality}

The energy of a cavity photon is determined by its longitudinal ($k_{z}$)
and transversal wavenumber ($k_{r}$) as $E=(\hbar c/n)\sqrt{k_{z}^{2}+k_{r}^{2}}$,
where $n$ again denotes the index of refraction of the medium. Owing
to the curvature of the mirrors, the boundary conditions for the photon
field depend on the distance to the optical axis $r=|\mathbf{r}|$.
For the longitudinal component, we set $k_{z}(\mathbf{r})=q\pi/D(r)$
where $q$ denotes the longitudinal mode number and $D(r)$ describes
the mirror separation as a function of $r$. For a symmetric resonator
consisting of two spherically curved mirror with separation $D_{0}$
and radius of curvature $R$, in a paraxial approximation ($r\ll R$,
$k_{r}\ll k_{z}$), the photon energy is given by \cite{Klaers:2010p2137}
\begin{equation}
E\simeq m_{\textrm{ph}}(c/n)^{2}+\frac{(\hbar k_{r})^{2}}{2m_{\textrm{ph}}}+\frac{1}{2}m_{\textrm{ph}}\Omega^{2}r^{2}\:,\label{eq:E}
\end{equation}
with an effective photon mass $m_{\textrm{ph}}=\pi\hbar nq/cD_{0}$
and trapping frequency $\Omega=c/n\sqrt{D_{0}R/2}$. This describes
a particle moving in the two-dimensional transversal plane with non-vanishing
(effective) mass subject to a harmonic trapping potential with trapping
frequency $\Omega$. Such a system is known to undergo Bose-Einstein
condensation at finite temperature \cite{Bagnato:1991p552}. If we
account for the two-fold polarization degeneracy of photons, condensation
is expected, when the particle number exceeds the critical particle
number
\begin{equation}
N_{\textrm{c}}=\frac{\pi^{2}}{3}\left(\frac{k_{\textrm{B}}T}{\hbar\Omega}\right)^{2}\;\textrm{.}\label{eq:Nc}
\end{equation}
The typical trapping frequency in our setup is $\Omega/2\pi\simeq41\,\textrm{GHz}$,
and at room temperature ($T=300\,\textrm{K}$) one obtains a critical
photon number of $N_{\textrm{c}}\simeq77,000$, which is experimentally
feasible. The physical reason for the possibility to observe Bose-Einstein
condensation at room temperature conditions is the small effective
photon mass $m_{\textrm{ph}}=\hbar\omega_{\textrm{c}}(n/c)^{2}\simeq7\cdot10^{-36}\,\textrm{kg}$,
which is ten orders of magnitude smaller than e.g. the mass of the
rubidium atom.

\subsection{Equilibrium versus non-equilibrium}

In general, particle loss can drive a system out of equilibrium, if
the timescale associated to loss is not well separated from the timescale
for the equilibration of the system. Separated timescales clearly
can be achieved for the case of dilute atomic gases. The true ground
state for e.g. an atomic rubidium gas is a cloud of molecular dimers.
However, researchers have learned in the 1980\textquoteright s to
the early 1990\textquoteright s that the recombination rate from three-body
collisions to the molecular state can be kept sufficiently small by
the use of very dilute atomic clouds for which the corresponding rates
are sufficiently small \cite{cornell_wieman_02,ketterle_02}. High
phase space densities can nevertheless be achieved by cooling to ultralow
temperatures in the nano-Kelvin regime. Correspondingly, quantum degeneracy
of a cloud of bosonic atoms can be reached under conditions that are
close to equilibrium. 

In the case of photons, non-equilibrium conditions can either arise
from a violation of the Kennard-Stepanov law (eq. \ref{eq:KS}) or
from a violation of chemical equilibrium (eq. \ref{eq:fugacity}),
if e.g. the photon loss rate is not negligible compared to the photon
absorption and emission rate. Clearly, the latter situation is well
known from typical laser operation. Both laser operation and Bose-Einstein
condensation, either of photons or atoms \cite{lee2000}, rely on
Bose-enhancement. However, to achieve lasing at the desired wavelength,
it is usually necessary to break the chemical equilibrium between
photons and molecules, allowing for a departure from Bose-Einstein
statistics and for a photon energy distribution independent of energetics.
For this purpose, gain and loss are deliberately engineered, for example,
by frequency-selective components. In the field of exciton-polaritons,
the question whether a system that is pumped and exhibits losses should
be regarded as polariton laser or polariton Bose-Einstein condensate
has been extensively discussed \cite{deng2006,Kasprzak:2008p716,wouters2008},
see also following articles in this volume. For the case of photonic
Bose-Einstein condensation, the role of losses and pumping has been
theoretically investigated by Kirton and Keeling \cite{Kirton:2013}.
Experimentally, the crossover between equilibrium and non-equilibrium
photon gases has been studied both in the non-degenerate and in the
quantum degenerate regime \cite{Klaers:2010p2137,marelic2014,schmitt2015}.

\section{Experiments on photon condensation}

A scheme of the setup used in the Bonn photon condensation experiment
is shown in Fig. \ref{fig:Figure1}a. The optical resonator consists
of two highly reflecting spherically curved mirrors ($\simeq0.999985$
reflectivity in the relevant wavelength region) with radius of curvature
$R=1\,\textrm{m}$. One of mirrors is cut to $\simeq1\,\textrm{mm}$
surface diameter to allow for a cavity length in the micrometer range
($D_{0}\simeq1.46\,\mu\textrm{m}$), as measured by the cavity free
spectral range, despite the mirror curvature. The resonator contains
a drop of liquid dye, typically rhodamine 6G or perylenedimide (PDI),
solved in an organic solvent. Both of these dyes have high quantum
efficiencies between 0.95 and 0.97, and fulfill the Kennard-Stepanov
relation in good approximation. Fig. \ref{fig:Figure1}b shows the
cavity spectrum (top) along with the absorption and fluorescence spectrum
for rhodamine dye (bottom). The resonator setup is off-resonantly
pumped with a laser beam near 532 nm wavelength derived from a frequency
doubled Nd:YAG laser inclined under 45$^{\circ}$ angle to the cavity
axis.

In initial experiments, the thermalization of the two-dimensional
photon gas in the dye-filled microresonator was carefully tested \cite{Klaers:2010p2137}.
Fig. 2a shows experimental spectra of the light transmitted through
one cavity mirror for two different temperatures of the setup (top:
$T\simeq300\,\textrm{K}$, room temperature; bottom $T\simeq365\,\textrm{K}$).
In these experiments, the average photon number inside the cavity
($N\simeq50$) is three orders of magnitude below the critical particle
number. The experimental data (dots) in both cases is well described
by a Boltzmann distribution of photon energies at the corresponding
temperature (solid line). In other experiments, the pump spot was
transversely displaced by a variable amount and the position where
the maximum of the observed fluorescence occurs was monitored, see
Fig. 2b. As expected in the presence of a trapping potential, a spatial
relaxation of the photons towards regions of low potential energy
near the optical axis was observed.
\begin{figure}
\begin{centering}
\includegraphics[width=0.9\columnwidth]{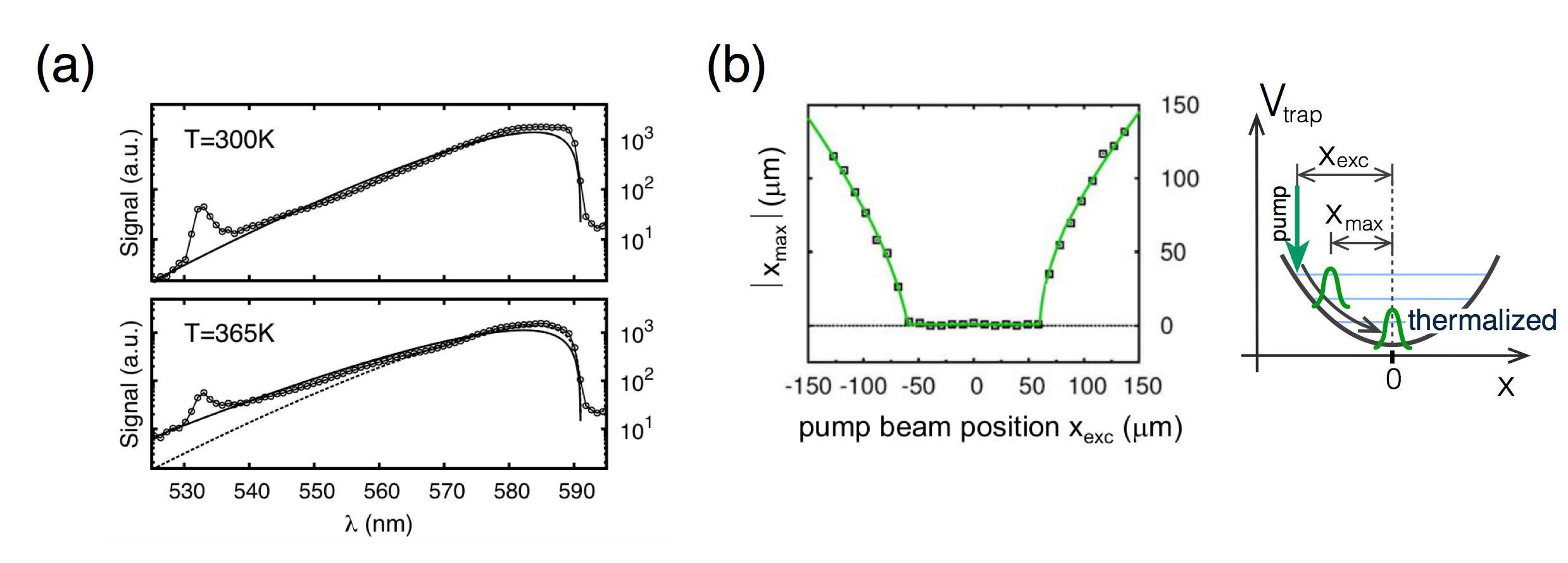}
\par\end{centering}

\centering{}%
\begin{minipage}[t]{0.93\columnwidth}%
\textbf{\small{}Figure 2 |}{\small{} }\textbf{\small{}(a)}{\small{}
Measured spectral intensity distributions (connected dots) of the
cavity emission for temperatures of the resonator setup of $300\,\textrm{K}$
(top) and $365\,\textrm{K}$ (bottom) at an average photon number
of $60\pm10$ inside the cavity, i.e. far below the onset of a BEC.
The solid lines are theoretical spectra based on a Bose-Einstein distribution.
For illustration a $T=300\,\textrm{K}$ distribution is also inserted
in the bottom graph (dashed line). }\textbf{\small{}(b)}{\small{}
Distance of the fluorescence intensity maximum from the optical axis
$\left|x_{\textrm{max}}\right|$ versus transverse position of the
pump spot, $x_{\textrm{exc}}$. Due to the thermalization, the photon
gas accumulates in the trap center, where the potential exhibits a
minimum value. This holds as long as the excitation spot is closer
than approximately $60\,\mu\textrm{m}$ distance. Figure taken from
Ref. \cite{Klaers:2010p2137}.}%
\end{minipage}
\end{figure}

\begin{figure}
\begin{centering}
\includegraphics[width=0.88\columnwidth]{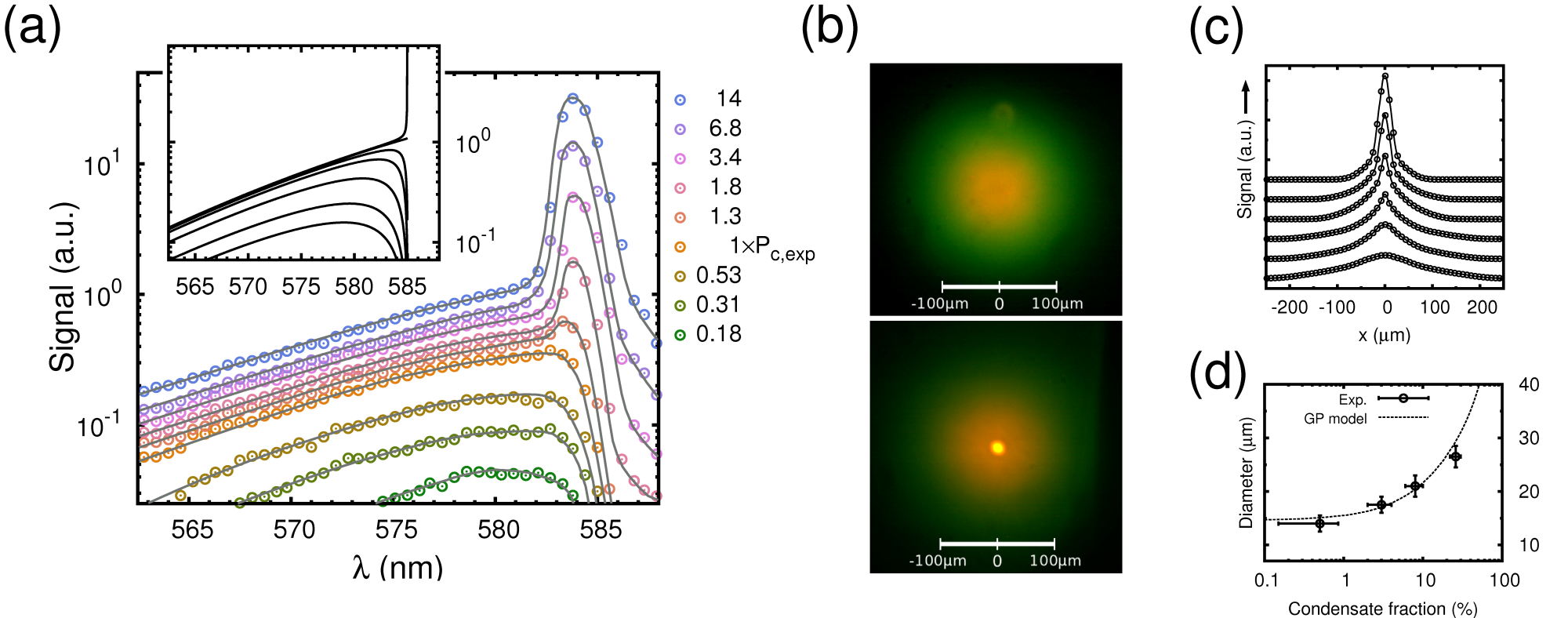}
\par\end{centering}

\centering{}%
\begin{minipage}[t]{0.93\columnwidth}%
\textbf{\small{}Figure 3 |}{\small{}\label{fig:Figure3}}\textbf{\small{}
(a)}{\small{} The connected circles show measured spectral intensity
distributions for different pump powers. The legend gives the optical
intracavity power, determining the photon number. On top of a broad
thermal wing, a spectrally sharp condensate peak at the position of
the cavity cutoff is visible above a critical power. The observed
peak width is limited by the spectrometer resolution. The inset gives
theoretical spectra based on Bose-Einstein distributed transversal
excitations. }\textbf{\small{}(b)}{\small{} Images of the radiation
emitted along the cavity axis, below (top) and above (bottom) the
critical power. In the latter case, a condensate peak is visible in
the center. }\textbf{\small{}(c)}{\small{} Cuts through the center
of the observed intensity distribution for increasing optical pump
powers. }\textbf{\small{}(d)}{\small{} The data points give the measured
width of the condensate peak versus condensate fraction and the dotted
line is the result of a theoretical model based on the Gross-Pitaevskii
equation. Figure taken from Ref. \cite{Klaers:2010}.}%
\end{minipage}
\end{figure}

In subsequent experiments, the dye-filled microcavity was operated
at photon numbers sufficiently high to reach quantum degeneracy. To
avoid excessive population of dye molecules in triplet states and
heat deposition, the optical pump beam was acousto-optically chopped
to $0.5\,\mu\textrm{s}$ long pulses, with a $8\,\textrm{ms}$ repetition
time. Fig. \ref{fig:Figure3}a shows typical spectra of the photon
gas at different photon numbers \cite{Klaers:2010}.While the observed
spectrum resembles a Boltzmann-distribution at small intracavity optical
powers, near the phase transition a shift of the maximum towards the
cutoff frequency is observed, and the spectrum more resembles a Bose-Einstein
distribution. At intracavity powers above the critical value, the
Bose-Einstein condensate occurs as a spectrally sharp peak at the
position of the cutoff. The observed spectral width of the condensate
peak is limited by the resolution of the used spectrometer. The experimental
results are in good agreement with theoretical expectations (see the
inset of the figure). At the phase transition, the optical intracavity
power is $P_{\textrm{c},\textrm{exp}}=(1.55\pm0.6)\,\textrm{W}$,
which corresponds to a photon number of $(6.3\pm2.4)\cdot10^{4}$.

Fig. \ref{fig:Figure3}b shows spatial images of the light transmitted
through one of the cavity mirrors (real image onto a color CCD camera)
both below (top) and above (bottom) the condensate threshold. Both
images show a shift from the yellow spectral regime for the transversally
low excited cavity modes located near the trap center to the green
for transversally higher excited modes appearing at the outer trap
regions. In the lower image, a bright spot is visible in the center
with a measured FWHM diameter of $(14\pm2)\,\mu\textrm{m}$. Within
the quoted experimental uncertainties, this corresponds well to the
expected diameter of the $\textrm{TEM}_{00}$ transverse ground state
mode of $12.2\,\mu\textrm{m}$, yielding clear evidence for a single-mode
macroscopic population of the ground state. Fig. \ref{fig:Figure3}c
gives normalized intensity profiles (cuts along one axis through the
trap center) for different powers. One observes that not only the
height of the condensate peak increases for larger condensate fractions,
but also its width, see also Fig. \ref{fig:Figure3}d. This effect
is not expected for an ideal photon gas, and suggests a weak repulsive
self-interaction mediated by the dye solution. The origin of the self-interaction
is thermal lensing, which under steady state conditions can be described
by a non-linear term analogous to the Gross-Pitaevskii equation (see
\cite{Klaers:2010}). In general, the interplay between between optical
and heat flow equations can lead to non-local interactions, see \cite{strinati2014}.
By comparing the observed increase of the mode diameter with numerical
solutions of the two-dimensional Gross-Pitaevskii equation, a dimensionless
interaction parameter of was estimated \cite{Klaers:2010}. This interaction
parameter is found to be significantly smaller than the values reported
for two-dimensional atomic physics quantum gas experiments \cite{Hadzibabic:2006,Clade:PRL2009}
and also below the values at which Kosterlitz-Thouless physics can
be expected to become important in the harmonically trapped case \cite{Hadzibabic:2009p2114}.
Experimentally, when directing the condensate through a Michelson-type
sheering interferometry, no signatures of phase blurring (that occur
in two-dimensional atomic gas experiments) were observed \cite{Klaers:2011}. 

Further signatures consistent within the framework of Bose-Einstein
condensation include the expected scaling of the critical photon number
with resonator geometry, and a spatial relaxation process that leads
to a strongly populated ground mode even for a spatially displaced
pump spot \cite{Klaers:2010}.

\section{\label{sec:section_fluctuations}Fluctuations of photon condensates}

\subsection{\label{sub:Fluctuation-map-of}Photon condensates coupled to a particle
reservoir}

In this section, we discuss quantum statistical properties of photon
condensates, in particular the photon number distribution and particle
number fluctuations. The main result is that photon Bose-Einstein
condensates in the dye microcavity system, owing to the grand-canonical
nature of the light-matter thermalization process, can show unusually
large particle number fluctuations, which are not observed in present
atomic Bose-Einstein condensates.

In statistical physics, different statistical ensembles reflect different
laws of conservation that can be realized in experiments. The micro-canonical
ensemble corresponds to a physical system with energy and particle
number strictly fixed at all times, while in the canonical ensemble
energy fluctuates around a mean value determined by the temperature
of a heat reservoir. Under grand-canonical conditions, both an exchange
of energy and particles with a large reservoir is allowed leading
to fluctuations in both quantities. The here investigated photon gas
in the dye microcavity, with photons being frequently absorbed and
emitted by dye molecules, belongs the latter class of experiments.
As discussed in section 2, absorption and emission can be regarded
as the two directions of a photochemical reaction $\gamma\,+\,\downarrow\,\rightleftarrows\,\uparrow$,
where photons ($\gamma$), ground state ($\downarrow$) and excited
dye molecules ($\uparrow$) are repeatedly converted into each other,
and the dye molecules act as a \textquotedblleft reservoir species\textquotedblright{}
for the photon gas.

A common assumption is that the ensemble conditions realized in a
physical system are not essential for its physical behavior. The various
statistical approaches are correspondingly expected to become interchangeable
in the thermodynamic limit \cite{Huang:StatisticalMechanics1987,Huang:Introduction2001},
in the sense that relative fluctuations vanish in all of them, i.e.
$\delta N/N\rightarrow0$ for the average total particle number $N$
and its root mean square deviation $\delta N$. This assumption is
however violated in the grand-canonical treatment of the ideal Bose
gas, where the occupation of any single particle state undergoes relative
fluctuations of 100\% of the average value \cite{ZIFF:1977p513,Kocharovsky:2006p985}.
For a macroscopically occupied ground state of a Bose-Einstein condensed
gas, this implies fluctuations of order of the total particle number,
i.e. $\delta N\simeq N$. While one usually expects fluctuations to
freeze out at low temperatures, here the reverse situation is encountered:
the total particle number starts to strongly fluctuate as the condensate
fraction approaches unity, a behavior that has been recognized early
in BEC theory \cite{Fierz:1955} and later has been termed ``grand-canonical
fluctuation catastrophe'' \cite{_Grossmann:1996,Holthaus:1998p198,Kocharovsky:2006p985}.
In experiments with cold atoms, this anomaly has not been observed
so far, as sufficiently large particle reservoirs are usually not
experimentally realizable. For those systems, much theoretical work
has been performed to obtain the particle number fluctuations in a
(micro-)canonical description \cite{Fujiwara:1970p993,_Grossmann:1996,_Politzer:1996,_Navez:1997},
and accounting for trapping potentials \cite{_Grossmann:97,_Weiss:97}.
A review can be found in reference \cite{Kocharovsky:2006p985}. Noteworthy,
the micro-canonical ensemble description of the ideal Bose gas shows
interesting connections to the partioning and factorizing problem
of integer numbers \cite{Weiss2004586}.

For a photon Bose-Einstein condensate, grand-canonical ensemble conditions
can be an inherent feature of the thermalization process and can therefore
influence the second order coherence properties \cite{Klaers2:2011}.
We consider a situation in which the photon condensate is coupled
to the electronic transitions of $M$ dye molecules (located in the
volume of the electromagnetic ground mode) by absorption and emissions
processes. In this way, the condensate exchanges excitations with
a reservoir of a given (finite) size. Using a master equation approach
one can show that the probability $\mathcal{P}_{n}$ to find $n$
photons in the ground state follows
\begin{eqnarray}
\frac{\mathcal{P}_{n}}{\mathcal{P}_{0}} & = & \frac{\left(M-X\right)!\,X!}{\left(M-X+n\right)!\,\left(X-n\right)!}\,e^{-n\hbar(\omega_{\textrm{c}}-\omega_{\textrm{ZPL}})/k_{\textrm{B}}T}\;\textrm{,}\label{eq:Pn}
\end{eqnarray}
where the excitation number $X$ is defined as the sum of ground mode
photon number and electronically excited molecules in the reservoir.
As before, $\omega_{\textrm{c}}$ and $\omega_{\textrm{ZPL}}$ denote
the frequencies of the condensate mode and zero-phonon-line of the
medium, respectively. In this calculation, $X$ is constant, i.e.
it is not expected to perform large fluctuations on its own. The photon
number distribution, which can also be derived in a superstatistical
approach \cite{Sobyanin:2012}, in general interpolates between Bose-Einstein
statistics and Poisson statistics. Assuming that the excitation level
$\rho_{\uparrow}/\rho_{\downarrow}\simeq X/(M-X)$ of the medium stays
fixed, which conserves the chemical potential $\mu$ (eq. \ref{eq:fugacity}),
the average condensate number $\bar{n}_{0}$ and total particle number
$\bar{N}$, one finds that large reservoirs $M$ lead to Bose-Einstein-like
statistics with an exponentially decaying photon number distribution
starting at a maximum value for zero photon number $n=0$. For small
reservoirs $\mathcal{P}_{n}$ becomes poissonian with a maximum value
at a non-zero photon number. The distinction between these two statistical
regimes is not unambiguous due to the smooth crossover behavior between
them. A natural choice for a borderline is the point at which 'finding
zero photons' ceases to be the most probable event, which occurs at
$\mathcal{P}_{0}=\mathcal{P}_{1}$ and resembles a common laser threshold
definition \cite{Scully:1967p208}. For the temperature $T_{x}$ at
which this condition is reached, given a certain system size $\bar{N}$
(average photon number) and the reservoir size $M$, one obtains the
equation 
\begin{equation}
\bar{N}-\frac{\pi}{6}\left(\frac{k_{B}T_{\text{x}}}{\hbar\Omega}\right)^{2}\simeq\sqrt{\frac{M/2}{1\hspace{-0.5mm}+\cosh\frac{\hbar\Delta}{k_{\textrm{B}}T_{\text{x}}}}}\enskip\textrm{.}\label{eq:Tx}
\end{equation}
Here $\Delta$ denotes the detuning between condensate mode and zero-phonon-line
of the dye, defined as $\Delta=\omega_{\textrm{c}}-\omega_{\text{ZPL}}$.
For zero dye-cavity detuning $\Delta=0$, one finds the analytic solution
$T_{\textrm{x},\Delta=0}\simeq T_{\textrm{c}}\,\sqrt{1-\sqrt{M}/2\bar{N}}$,
provided that $\sqrt{M}/2\bar{N}<1$. For general detunings $\Delta$,
equation (\ref{eq:Tx}) has to be solved numerically. 
\begin{figure}
\begin{centering}
\includegraphics[width=0.45\paperwidth]{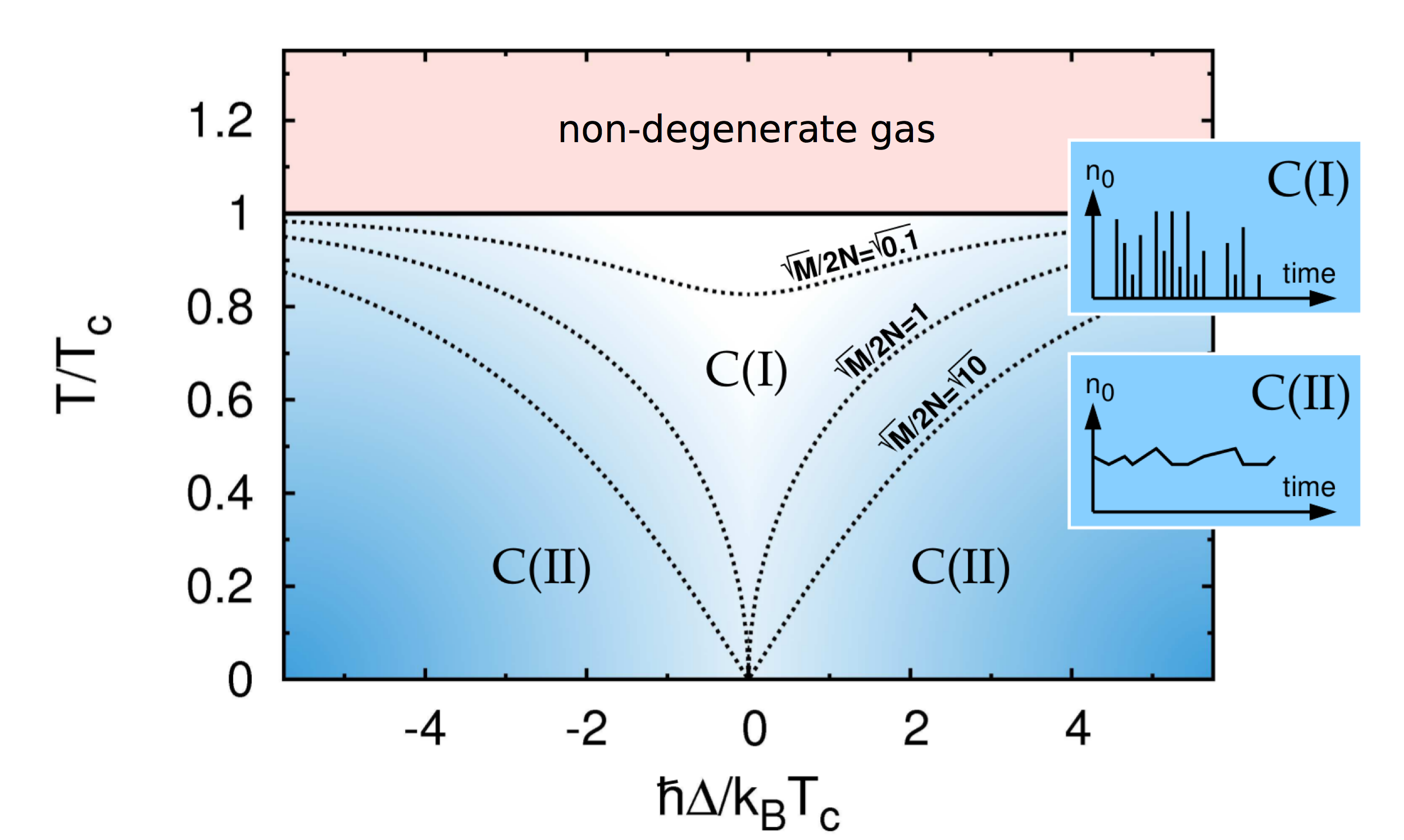}
\par\end{centering}

\centering{}%
\begin{minipage}[t]{0.93\columnwidth}%
\textbf{\small{}Figure 4 |}{\small{}\label{fig:Figure4} Phase diagram
of the two-dimensional photon gas for fixed average photon number
$\bar{N}$ in the plane spanned by the reduced temperature $T/T_{\textrm{c}}$
and the dye-cavity detuning $\hbar\Delta/k_{\textrm{B}}T_{\textrm{c}}$.
The solid line marks the BEC phase transition. The dashed lines (three
cases are shown) separate two regimes: a condensate regime with large
number fluctuations and a Bose-Einstein-like photon number distribution
C(I), and a regime of non-fluctuating condensates obeying Poisson
statistics C(II). The temperature of the crossover C(I)-C(II) depends
on the ratio $\sqrt{M}/\bar{N}$, where the reservoir size $M$ denotes
the number of dye molecules in the mode volume of the ground state.
The insets give a sketch of the corresponding temporal evolution of
the condensate photon number $n_{0}(t)$. Figure taken from Ref. \cite{Klaers2:2011}.}%
\end{minipage}
\end{figure}
Figure (\ref{fig:Figure4}) gives a phase diagram, where solutions
for three different cases $\sqrt{M}/2\bar{N}=\sqrt{0.1},1,\sqrt{10}$
are marked as dashed lines, which separates two different regimes
of the photon condensate, denoted by C(I) with Bose-Einstein-like
photon statistics and C(II) with Poisson statistics, respectively.
In terms of second order correlations, the dashed lines correspond
to $g^{(2)}(0)\simeq\pi/2$, or relative fluctuations of $\delta n/\bar{n}_{0}=\sqrt{g^{(2)}(0)-1}=0.75$.
Note that both $T_{\text{c}}$ and $T_{\text{x}}$ are conserved in
a thermodynamic limit $\bar{N},\,M,\,R\rightarrow\infty$ in which
$\bar{N}/R=\text{const}$ and $\sqrt{M}/\bar{N}=\text{const}$. A
recent theory work has investigated the possible effects of fast photon-photon
interactions on the photon number statistics \cite{vanderWurff2014}.

\subsection{Observation of anomalous condensate fluctuations}

The intensity correlations and fluctuations of the condensate have
been measured using a Hanbury Brown-Twiss setup \cite{Schmitt:PhysRevLett.112.030401,Ciuti:Physics.7.7}.
In this experiment, the condensate mode is separated from the higher
transversal modes by spatial filtering in the far field, which corresponds
to a transverse momentum filter. The beam is split into two paths,
each of which are directed onto single-photon avalanche photodiodes.
Time correlations of the condensate population can be determined with
a temporal resolution of 60ps with this setup. The second-order correlation
function $g^{(2)}(t_{1},t_{2})=\left\langle n_{0}(t_{1})n_{0}(t_{2})\right\rangle /\left\langle n_{0}(t_{1})\right\rangle \left\langle n_{0}(t_{2})\right\rangle $
to good approximation is found to depend only on the time delay $\tau=t_{2}-t_{1}$.
Further analysis is thus performed with the time averaged function
$g^{(2)}(\tau)=\left\langle g^{(2)}(t_{1},t_{2})\right\rangle _{\tau=t_{2}-t_{1}}$. 

\begin{figure}
\begin{centering}
\includegraphics[width=0.38\paperwidth]{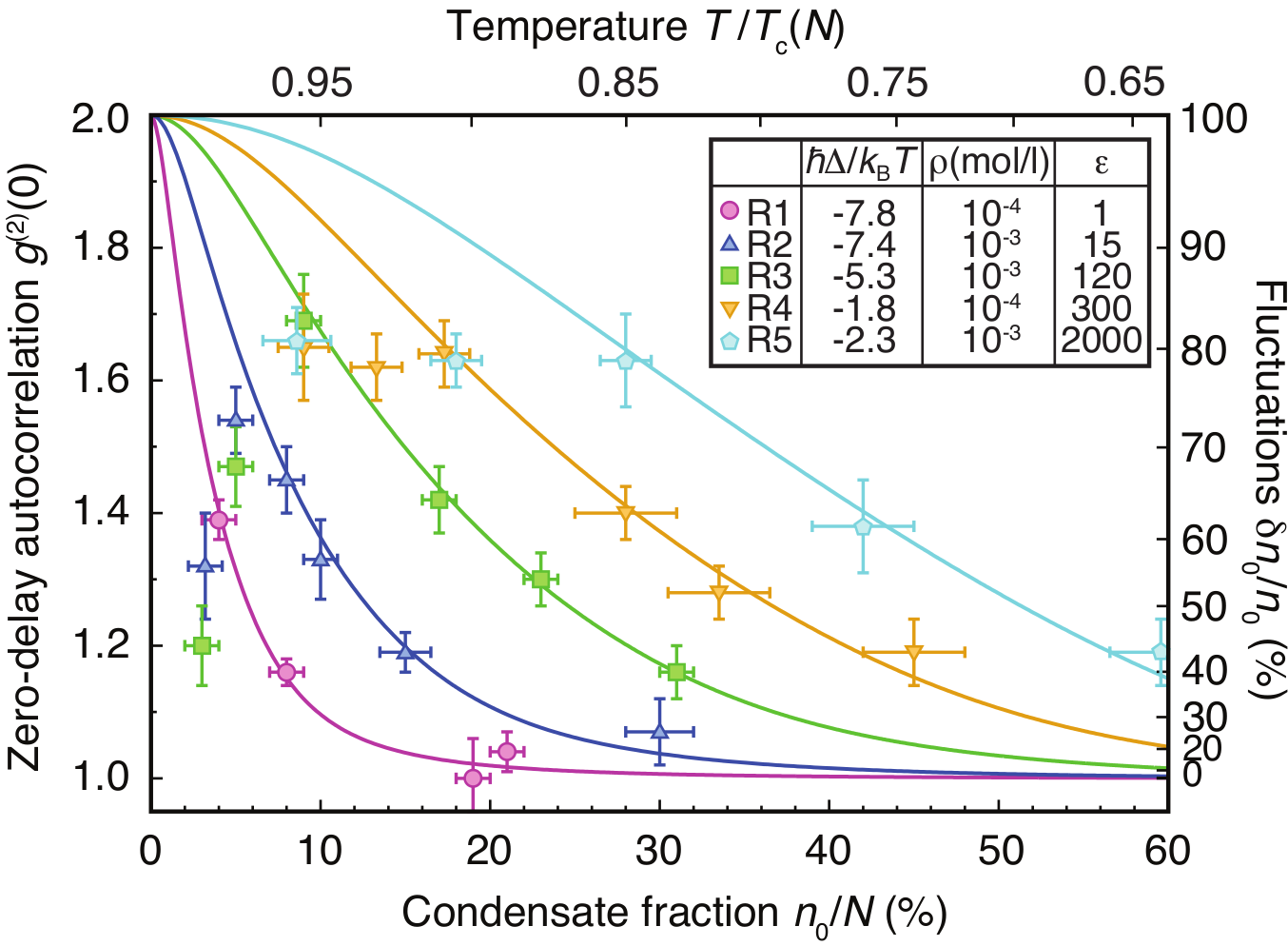}
\par\end{centering}

\centering{}%
\begin{minipage}[t]{0.93\columnwidth}%
\textbf{\small{}Figure 5 |}{\small{}\label{fig:Figure5} Zero-delay
autocorrelations $g^{(2)}(0)$ and condensate fluctuations $\delta n_{0}/\bar{n}_{0}$
versus condensate fraction $\bar{n}_{0}/\bar{N}$ {[}or equivalently
the corresponding reduced temperature $T/T_{\text{c}}(\bar{N})$ at
$T=300\,\text{K}${]}, for five different reservoirs R1-R5. The increase
of the effective molecular reservoir size from R1 to R5 is quantified
by the parameter $\epsilon$ (third column), defined in equation (\ref{eq:epsilon}).
Condensate fluctuations extend deep into the condensed phase for high
dye concentration $\rho$ and small dye-cavity detuning $\Delta$
(R5). Results of a theoretical model based on equation (\ref{eq:Pn})
are shown as solid lines. The error bars indicate statistical uncertainties.
Experimental parameters: condensate wavelength $\lambda_{0}=\{598,\,595,\,580,\,598,\,602\}\,\text{nm}$
for data sets R1-R5; dye concentration $\rho=\{10^{-4},\,10^{-3},\,10^{-3}\}\,\text{mol}/l$
for R1-R3 (rhodamine 6G), and $\rho=\{10^{-4},\,10^{-3}\}\,\text{mol}/l$
for R4-R5 (perylene red). For the theory curves, we find effective
reservoir sized of $M=\{5.5\pm2.2,\,20\pm7,\,16\pm6,\,2.1\pm0.4,\,11\pm4\}\times10^{9}$
for R1-R5. Figure taken from Ref. \cite{Schmitt:PhysRevLett.112.030401}.}%
\end{minipage}
\end{figure}

We have varied the reservoir size systematically to test for the grand-canonical
nature of the system. Figure (\ref{fig:Figure5}) shows the zero-delay
correlations $g^{(2)}(0)$ as a function of the condensate fraction
$\bar{n}_{0}/\bar{N}$ for five combinations of dye concentration
$\rho$ and dye-cavity detuning $\Delta$.
 The data sets labelled with R1-R3 have been obtained with rhodamine
6G dye ($\omega_{\text{ZPL}}=2\pi c/545\text{nm}$). For measurements
R4 and R5, we have used perylene red ($\omega_{\text{ZPL}}=2\pi c/585\text{nm}$)
as dye species, which allows us to reduce the detuning between condensate
and dye reservoir and to effectively increase the reservoir size.
Following equation (\ref{eq:Tx}), this effective reservoir size can
be quantified as
\begin{equation}
M_{\text{eff}}=\frac{M/2}{1\hspace{-0.5mm}+\cosh(\hbar\Delta/k_{\textrm{B}}T)}\;\text{.}\label{eq:Meff}
\end{equation}
Furthermore, a relative reservoir size is obtained by normalizing
to the reservoir size in measurement R1
\begin{equation}
\epsilon=\frac{M_{\text{eff},R_{i}}}{M_{\text{eff},R1}}=\frac{\rho_{R_{i}}}{\rho_{R1}}\times\frac{1+\cosh(\hbar\Delta_{R1}/k_{\text{B}}T)}{1+\cosh(\hbar\Delta_{R_{i}}/k_{\text{B}}T)}\;\text{.}\label{eq:epsilon}
\end{equation}
For the lowest dye concentration and largest detuning (R1, $\epsilon=1$),
the particle reservoir is so small that the condensate fluctuations
are damped almost directly above the condensation threshold ($\bar{N}\ge N_{\text{c}}$).
By increasing dye concentration and decreasing the dye-cavity detuning
one can systematically extend the regime of large fluctuations to
higher condensate fractions (R1-R5). Our experimental results are
recovered by a theoretical modeling shown as solid lines in figure
(\ref{fig:Figure5}), except for small condensate fractions below
5\%. The here visible drop-off in the correlation signal is attributed
to imperfect mode filtering, which does not fully preclude photons
in higher transversal cavity modes that are statistically uncorrelated
to the ground mode photons from reaching the avalanche photo detectors.
The maximum observed zero-delay autocorrelation is $g^{(2)}(0)\simeq1.67$,
corresponding to relative fluctuations of $\delta n_{0}/\bar{n}_{0}=82\%$,
which is slightly less than theoretically expected. For the largest
reservoir realized (R5, $\epsilon=2000$), we observe zero-delay correlations
of $g^{(2)}(0)\simeq1.2$ at a condensate fraction of $\bar{n}_{0}/\bar{N}\simeq0.6$.
At this point, the condensate still performs large relative fluctuations
of $\delta n_{0}/\bar{n}_{0}=\sqrt{g^{(2)}(0)-1}\simeq45\%$, although
its occupation number is comparable to the total photon number. This
clearly demonstrates that the observed super-Poissonian photon statistics
is determined by the grand-canonical particle exchange between condensate
and dye reservoir.

\section{Conclusions}

We have described recent experiments on photon Bose-Einstein condensation
in a dye-filled optical microresonator. Thermalization of the photon
gas is achieved by a fluorescence induced thermalization mechanism,
which establishes a thermal contact to the room temperature dye medium,
and allows for a freely adjustable chemical potential. The photons
here act like a gas of material particles with a phase transition
temperature that is many orders of magnitude higher than for dilute
atomic Bose-Einstein condensates. A further notable system property
is a regime with unconventional fluctuation properties, in which statistical
fluctuations of the condensate number comparable to the total particle
number occur. This is a yet unexplored regime of Bose-Einstein condensation
that originates from the grand-canonical nature of the light-matter
thermalization process and breaks the usual assumption of ensemble
equivalence in statistical physics. Moreover, the unconventional second
order coherence properties of a photon condensate can draw a further
borderline (in addition to the equilibrated system state) to laser-like
behavior, if one follows the usual definition of a laser as a both
first and second order coherent light source.

For the future, it will be interesting to test for the first order
coherence of photon condensation in the grand-canonical limit, and
to verify whether such a condensate exhibits superfluidity. A further
fascinating perspective is the exploration of periodic potentials
for the photon gas, which may allow to tailor novel quantum manybody
states of light.

We acknowledge funding from the ERC (INPEC) and the DFG (We 1748-17). 

\ 

\begin{spacing}{0.7}
\noindent {\small{}\bibliographystyle{amsplain-thesis}
\bibliography{Klaers}
}{\small \par}\end{spacing}

\end{document}